# Afterglow from GRB 070610/Swift J195509.6+261406: an explanation using the fireball model

KONG SiWei & HUANG YongFeng[†]

Department of Astronomy, Nanjing University, Nanjing 210093, China

**GRB 070610, which is also named Swift J195509.6+261406, is a peculiar Galactic transient with significant variability on short timescales in both X-ray and optical light curves. One possible explanation is that GRB 070610/Swift J195509.6 + 261406 is a soft gamma-ray repeater (SGR) in our Galaxy. Here we use the fireball model which is usually recognized as the standard model of gamma-ray burst (GRB) afterglows, and the energy injection hypothesis to interpret the X-ray and optical afterglow light curves of GRB 070610/Swift J195509.6 + 261406. It is found that the model is generally consistent with observations.**

gamma ray bursts, soft gamma ray repeaters

## 1　Introduction

GRB 070610 was detected at 20:52:26 UT on 2007 June 10 by the Burst Alert Telescope (BAT) onboard the *Swift* satellite[1] and was located at the position of $\alpha = 19^h 55^m 13.1^s$, $\delta = +26° 15' 20"$ (J2000.0), with a 90% containment radius of 1.8'[2]. This location is just 1 degree away from the Galactic plane. The BAT data shows that this event is consisted of a single peak, with the duration being $T_{90} = 4.6 \pm 0.4$ s[2].

Due to an Earth limb constraint, the satellite did not slew to the BAT position promptly and the X-Ray Telescope (XRT) began observing the GRB 070610 field 3.2 ks after the BAT trigger[3]. A variable point-like source was detected by XRT in the BAT error circle[3]. The behavior of the XRT light curve is obviously different from that of a typical gamma-ray burst (GRB) X-ray afterglow. The flux of the XRT light curve is almost constant and shows no obvious descent until very late time. Furthermore, there are many flares in the XRT light curve. The most dramatic flare appears very late ($t \sim 7.86 \times 10^4$ s) and brightens by a factor of ~ 100 during a timescale of $\Delta t/t \sim 10^{-4}$[4].

Optical observations in the BAT error circle had also been done by various astronomers[5-8]. A new variable source was found at the location of the burst. The overall optical light curve shows strong flaring activities[4, 6, 9, 10] and is very different from the optical afterglow of a typical GRB.

These unusual behaviors in both the X-ray and optical light curves, together with the fact that the location of the burst is in the Galactic plane, suggest that GRB

Received May 20, 2009; accepted Jun 03, 2009
doi:
[†]Corresponding author (email: hyf@nju.edu.cn)
Supported by the National Natural Science Foundation of China (Grant 10625313), and by National Basic Research Program of China (973 Program 2009CB824800).

070610 may not be a classical GRB, but should be a Galactic transient[7]. So GRB 070610 is re-named as Swift J195509.6 + 261406.

One possible explanation is that Swift J195509.6 + 261406 is a soft gamma-ray repeater (SGR)[4, 9, 10]. SGRs are a special type of γ-ray transient sources with soft spectra and unpredictable recurrences[11]. They are thought to be produced by young neutron stars with extremely high magnetic fields, which are named magnetars[12]. A typical SGR burst emits a huge amount of energy of about $10^{39} - 10^{41}$ ergs in just several hundred milliseconds. Similar to GRBs, such a huge energy release within a small volume and on the short timescale will inevitably produces a fireball. So the fireball model may be a reasonable explanation for the afterglow of SGR bursts[13].

In this paper, we use the fireball model to reproduce the unusual X-ray and optical afterglow light curves of GRB 070610/Swift J195509.6 + 261406. The outline of our paper is as follows. We introduce our model in Section 2, and present our numerical results and the fit to observations in Section 3. Section 4 presents our conclusions and discussion.

## 2 Model

In the standard fireball model, the outflow of a GRB, which moves relativistically, interacts with the surrounding medium to form an external shock. A constant fraction $\varepsilon_e$ of the shock energy will be transferred to the swept-up electrons and accelerate them to relativistic velocities. Similarly, a constant fraction $\varepsilon_B$ of the shock energy will go to the magnetic field. The shock-accelerated relativistic electrons move in the magnetic field and emit synchrotron radiation. The resulting spectrum and light curve can be approximated as simple broken power-law functions.

We use the convenient equations developed by Huang et al.[14−17] to describe the dynamics of the outflow. The evolution of the bulk Lorentz factor $\gamma$, the shock radius $R$, and the swept-up medium mass $m$, is described by three differential equations,

$$\frac{d\gamma}{dm} = -\frac{\gamma^2 - 1}{M_{ej} + \varepsilon m + 2(1-\varepsilon)\gamma m} \quad (1)$$

$$\frac{dR}{dt} = \beta c \gamma \left(\gamma + \sqrt{\gamma^2 - 1}\right) \quad (2)$$

$$\frac{dm}{dR} = 2\pi R^2 (1 - \cos\theta) n m_p \quad (3)$$

where $m_p$ is the mass of proton, $M_{ej}$ is the initial mass of the outflow, $\theta$ is the half opening angle of the jet, $n$ is the number density of the environment, and $\beta = \sqrt{\gamma^2 - 1}/\gamma$. $\varepsilon$ is the radiative efficiency, which equals 1 for the highly radiative case, and equals 0 in the adiabatic case. Since the ejecta becomes adiabatic several hours after the trigger, in this paper we assume the fireball is completely adiabatic all the time, so that $\varepsilon = 0$. In the simplest case, the interstellar medium (ISM) should be homogeneous. So, we take the number density $n$ of the circum-burst environment as a constant value.

In our calculations, we assume that the emission mainly comes from the synchrotron radiation of the shocked electrons. The equal arrival time surface (EATS) effect [18] is considered in our work. We ignore the sideways expansion of the jet, because many numerical simulations indicate that it is a very slow process[19−21].

The most attractive feature of the X-ray and optical light curves of GRB 070610/Swift J195509.6 + 261406 is the presence of many dramatic flares. As mentioned above, the light curve produced by a standard fireball is usually a smooth broken power-law function. So the standard fireball model itself can not describe these flaring behaviors. We suggest that these flares should be associated with some sudden energy injections of the central engine. Here, we will mainly concentrate on the overall afterglow behavior. For simplicity, we approximate those discrete energy injections as a continuous energy supply to the fireball that lasts for about two days. The injection power is assumed to take the form

$$\frac{dE_{inj}}{dt} = \bar{Q} t^q \quad \text{for} \quad t_{start} < t < t_{end}, \quad (4)$$

where $\bar{Q}$ and $q$ is constant, $t_{start}$ is the beginning time of the energy injection, and $t_{end}$ is the ending time of the energy injection. So the evolution of the bulk Lorentz factor $\gamma$ now should be calculated from

$$\frac{d\gamma}{dt} = \frac{1}{M_{ej} + \varepsilon m + 2(1-\varepsilon)\gamma m} \times \left(\frac{1}{c^2}\frac{dE_{inj}}{dt} - (\gamma^2 - 1)\frac{dm}{dt}\right). \quad (5)$$

## 3 Numerical results

In this section we use the model described above to reproduce the observed light curves of GRB 070610/Swift J195509.6 + 261406. In our calculation, we choose the parameters as $\theta = 0.12$ radian, the isotropic equivalent kinetic energy of the explosion $E_0 = 1.0 \times 10^{42}$ ergs, $\varepsilon_e = 0.13$, $\varepsilon_B = 0.017$, the electron energy distribution power-law index $p = 2.1$, $n = 1.0$ cm$^{-3}$, $\bar{Q} = 3.8 \times 10^{36}$, $q = 0$, $t_{start} = 3.2 \times 10^3$ s, $t_{end} = 2.0 \times 10^5$ s, and the luminosity distance of the source $D_L = 5.0$ kpc. The observed X-ray and optical light curves of GRB 070610/Swift J195509.6+261406 and our best fit are illustrated in Figures 1 and 2 respectively.

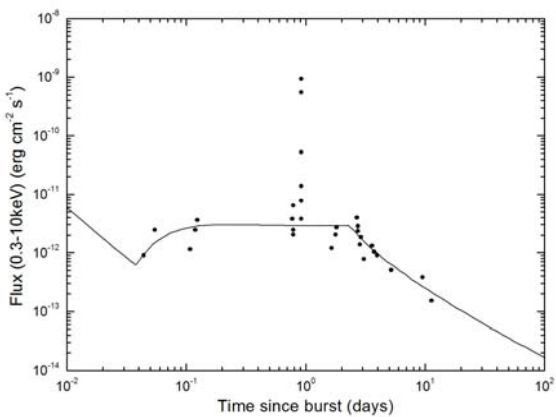

**Figure 1** Our best fit to the X-ray light curve of GRB 070610/Swift J195509.6 + 261406. The dots are observed data[10] and the solid line is our theoretical curve.

In Figure 1, we can see that there are many flares in the X-ray light curve of GRB 070610/Swift J195509.6 + 261406. One possible explanation is that these flares are associated with many separated and instantaneous energy injections from the central engine. The detailed mechanism for these flares is not the goal of this study, but a preliminary modeling can be found in Xu & Huang's work[22]. We have approximated these separated and instantaneous energy injections as one continuous and constant ($q = 0$) energy injection in our work to describe the general behavior of the light curve. It is clear that our model can reproduce the general features of the X-ray light curve very well. It suggests that the general afterglow could be produced by external shocks, just as in classical GRBs.

Figure 2 presents our theoretical I-band afterglow light curve for GRB 070610/Swift J195509.6 + 261406, together with the observed I-band flares. Note that observationally, people have only found flares in I-band, but not persistent optical afterglow. Our theoretical light curve is generally below the observed points, thus does not conflict with observations. In fact, the Galactic extinction in the direction of GRB 070610/Swift J195509.6 + 261406 is as large as $A_I = 6.384$ mag[23]. If the extinction is taken into account, then the theoretical lightcurve should be further shifted downward by 6.384 mag. This may explain why that people could only see the bright optical flares, but could not detect the persistent optical afterglow.

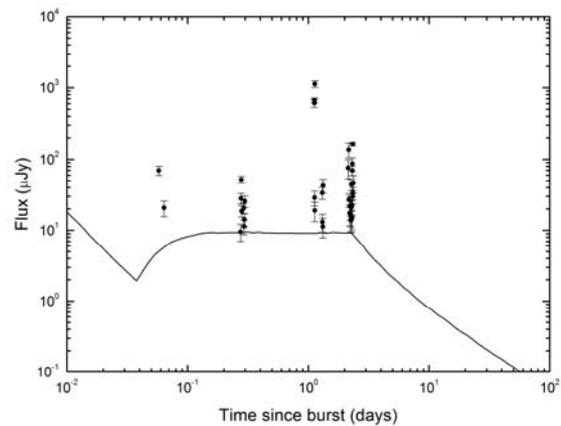

**Figure 2** Calculated I-band afterglow light curve of GRB 070610/Swift J195509.6 + 261406 and the observed optical flares. The dots are observed data[10] and the solid line is our theoretical curve.

## 4 Conclusion and discussion

GRB 070610/Swift J195509.6 + 261406 is a special transient. Many models have been suggested to explain the nature of the source[4, 9, 10, 22]. One possibility is that the source is a bursting pulsar like GRO J1744-28[24, 25]. But the lack of further gamma-ray detections is a difficulty for this possibility. Another explanation is that the source is a fast X-ray novae like the black hole candidate V4641 Sgr[26]. But again we lack the further radio and gamma-ray detections. The optical property of GRB 070610/Swift J195509.6+261406 is also different from that of V4641 Sgr.

The hypothesis that this source is a new SGR is also possible. Huang, Dai & Lu[13] have suggested that the fireball model can describe the general behaviors of SGR afterglows satisfactorily. For the curious flares in the light curves, we believe that they are associated with many separated and instantaneous energy injections from the central engine. These energy injections may come from the falling of some planet fragments onto the central compact star. The collision between these fragments and the central compact star will produce the flares via cyclotron radiation.[22] At the same time, a fraction of the collision energy will be injected into the external shock.

We have used the fireball model, together with a continuous energy injection process, to reproduce the observed light curves of GRB 070610/Swift J195509.6 + 261406 numerically. The isotropic equivalent kinetic energy of the explosion used in our modeling is $E_0 = 1.0 \times 10^{42}$ ergs. This is a typical value for SGR bursts. The total injected energy is about $7.5 \times 10^{41}$ ergs, almost equals to $E_0$. Other parameter values, such as $\theta$, $\varepsilon_e$, $\varepsilon_B$, $p$, are all typical in the standard fireball model.

If GRB 070610/Swift J195509.6 + 261406 is indeed a SGR, then it is the first SGR detected with optical flares. The multi-band observational data of SGRs will advance our studies on their properties. But the low Galactic latitude of this event prevents us from detecting any persistent optical afterglow. In the future, there will be more and more detailed and long time multi-band observations of SGRs. The nature of SGRs will be completely understood finally.

1  Pagani C, Barthelmy S D, Cummings J R, et al. GRB 070610: Swift detection of a burst. GCN Circ. 6489, 2007 (http://gcn.gsfc.nasa.gov/gcn/gcn3/6489.gcn3)
2  Tueller J, Barbier L, Barthelmy S D, et al. GRB 070610, Swift-BAT refined analysis. GCN Circ. 6491, 2007 (http://gcn.gsfc.nasa.gov/gcn/gcn3/6491.gcn3)
3  Pagani C, Kennea J A. GRB 070610: Swift-XRT position. GCN Circ. 6490, 2007 (http://gcn.gsfc.nasa.gov/gcn/gcn3/6490.gcn3)
4  Kasliwal M M, Cenko S B, Kulkarni S R, et al. GRB 070610: a curious Galactic transient. Astrophys J, 2008, 678: 1127-1135
5  Stefanescu A, Slowikowska A, Kanbach G, et al. GRB 070610: OPTIMA-Burst high-time-resolution optical observations. GCN Circ. 6492, 2007 (http://gcn.gsfc.nasa.gov/gcn/gcn3/6492.gcn3)
6  Postigo A D U, Castro-Tirado A J, Aceituno F. GRB 070610: Optical observations from OSN. GCN Circ. 6501, 2007 (http://gcn.gsfc.nasa.gov/gcn/gcn3/6501.gcn3)
7  Kann D A, Wilson A C, Schulze S, et al. GRB 070610: TLS RRM sees flaring behaviour - Galactic transient? GCN Circ. 6505, 2007 (http://gcn.gsfc.nasa.gov/gcn/gcn3/6505.gcn3)
8  Klotz A, Boer M, Atteia J L, et al. GRB 070610: TAROT Calern observatory optical observations. GCN Circ. 6513, 2007 (http://gcn.gsfc.nasa.gov/gcn/gcn3/6513.gcn3)
9  Stefanescu A, Kanbach G, Słowikowska A. et al. Very fast optical flaring from a possible new Galactic magnetar. Nature, 2008, 455: 503-505
10  Castro-Tirado A J, Postigo A D U, Gorosabel J, et al. Flares from a candidate Galactic magnetar suggest a missing link to dim isolated neutron stars. Nature, 2008, 455: 506-509
11  Norris J P, Hertz P, Wood K S, et al. On the nature of soft gamma repeaters. Astrophys J, 1991, 366: 240-252
12  Thompson C, Duncan R C. The soft gamma repeaters as very strongly magnetized neutron stars. II. Quiescent neutrino, X-ray, and Alfvén wave emission. Astrophys J, 1996, 473: 322-342
13  Huang Y F, Dai Z G, Lu T. Fireball/blastwave model and soft γ-ray burst. Chin Phys Lett, 1998, 15: 775-777
14  Huang Y F, Dai Z G, Lu T. A generic dynamical model of gamma-ray burst remnants. Mon Not Roy Astron Soc, 1999, 309: 513-516
15  Huang Y F, Dai Z G, Lu T. On the optical light curves of afterglows from jetted gamma-ray burst ejecta: effects of parameters. Mon Not Roy Astron Soc, 2000, 316: 943-949
16  Huang Y F, Gou L J, Dai Z G, et al. Overall evolution of jetted gamma-ray burst ejecta. Astrophys J, 2000, 543: 90-96
17  Huang Y F, Cheng K S. Gamma-ray bursts: optical afterglows in the deep Newtonian phase. Mon Not Roy Astron Soc, 2003, 341: 263-269
18  Huang Y F, Lu Y, Wong A Y L, et al. A detailed study on the equal arrival time surface effect in gamma-ray burst afterglows. Chin J Astron Astrophys, 2007, 7: 397-404
19  Granot J, Miller M, Piran T, et al. Light curves from an expanding relativistic jet. In: Costa E, Frontera F, Hjorth J, eds. Gamma-ray Bursts in the Afterglow Era. Berlin: Springer-Verlag, 2001. 312-314
20  Cannizzo J K, Gehrels N, Vishniac E T. A numerical gamma-ray burst simulation using three-dimensional relativistic hydrodynamics: the transition from spherical to jetlike expansion. Astrophys J, 2004, 601: 380-390
21  Zhang W Q, MacFadyen A. The dynamics and afterglow radiation of gamma-ray bursts. I. Constant Density Medium. Astrophys J, 2009, 698: 1261-1272
22  Xu M, Huang F Y. A model for the optical flares from the Galactic transient Swift J195509+261406. Sci China Ser G-Phys Mech Astron, 2009, in press (this volume)
23  Schlegel D J, Finkbeiner D P, Davis M. Maps of dust infrared emission for use in estimation of reddening and cosmic microwave background radiation foregrounds. Astrophys J, 1998, 500: 525-553
24  Kouveliotou C, van Paradijs J, Fishman G J, et al. A new type of transient high-energy source in the direction of the Galactic Centre. Nature, 1996, 379: 799-801
25  Sazonov S Y, Sunyaev R A, Lund N. Super-Eddington X-ray luminosity of the bursting pulsar GRO J1744-28: WATCH/GRANAT observations. Astron Lett, 1997, 23: 286-292
26  Revnivtsev M, Gilfanov M, Churazov E, et al. Super-Eddington outburst of V4641 Sgr. Astron Astrophys, 2002, 391: 1013-1022